\documentclass{aa}
\usepackage{graphicx}
\input{psfig.sty}

\def\es{ergs s$^{-1}\ $}

\def\lae{\mathrel{<\kern-1.0em\lower0.9ex\hbox{$\sim$}}}
\def\gae{\mathrel{>\kern-1.0em\lower0.9ex\hbox{$\sim$}}}

\begin{document} 


\title{A Hard Medium Survey with ASCA.
IV.: the Radio-Loud Type 2 QSO AX J0843+2942}

\author{R. Della Ceca \inst{1}, V. Braito \inst{1,2}, V. Beckmann \inst{3}, 
A. Caccianiga \inst{1}, I. Cagnoni \inst{4}, I.M. Gioia \inst{5}, T. Maccacaro \inst{1}, P. Severgnini \inst{1} 
and A. Wolter \inst{1}}

\offprints{R. Della Ceca}

\institute{
Osservatorio Astronomico di Brera, via Brera 28, I-20121 Milano, Italy. 
E-mail: rdc@brera.mi.astro.it, braito@brera.mi.astro.it, caccia@brera.mi.astro.it, 
tommaso@brera.mi.astro.it, paola@brera.mi.astro.it, anna@brera.mi.astro.it
\and
Dipartimento di Astronomia, Universit\`a degli Studi di Padova, vicolo dell'Osservatorio 2, 
I-35122, Padova, Italy 
\and
INTEGRAL Science Data Centre, Chemin d'Ecogia 16, CH-1290 Versoix, Switzerland. 
E-mail: Volker.Beckmann@obs.unige.ch
\and
Dipartimento di Scienze, Universit\`a dell'Insubria, Como, Italy, 
E-mail: ilaria.cagnoni@uninsubria.it
\and
Istituto di Radioastronomia CNR, via P Gobetti 101, I-40129,  Bologna, Italy
E-mail: gioia@ira.cnr.it
}

\date{Received 27 February 2003/ Accepted 26 May 2003}

\abstract{We discuss the X-ray, optical and radio properties of 
AX J0843+2942, a high luminosity Type 2 AGN 
found in the ASCA Hard Serendipitous Survey.  
The X-ray spectrum is best described by 
an absorbed power-law model with 
photon index of $\Gamma$ = $1.72^{+0.3}_{-0.6}$ and intrinsic 
absorbing column density of 
$N_H$ = $1.44^{+0.33}_{-0.52} \times 10^{23}$ cm$^{-2}$. 
The intrinsic luminosity in the 0.5 -- 10 keV energy band is   
$\simeq 3\times 10^{45}$ \es, well within the range of 
quasar luminosities.
AX J0843+2942, positionally coincident with the 
core of a triple and strong ($S_{1.4 GHz} \sim 1$ Jy; 
$P_{1.4 GHz} \sim 9\times 10^{33}$ erg s$^{-1}$ Hz$^{-1}$) 
radio source, 
is spectroscopically identified with a Narrow Line 
object (intrinsic FWHM of all the permitted emission lines 
$\lae 1200$ km s$^{-1}$) at z=0.398, having line features and 
ratios typical of Seyfert-2 like objects.
The high X-ray luminosity, coupled with the  high intrinsic
absorption, the optical spectral properties 
and the radio power, allow us to
propose AX J0843+2942 as a Radio-Loud ``Type 2 QSO".
A discussion of the SED of this object is presented here together with a  
comparison  with the SED of Ultra Luminous 
Infrared Galaxies, other ``Type 2 QSO" candidates from the literature, 
and ``normal" Radio-Quiet and Radio-Loud QSOs. 
}

\authorrunning{Della Ceca et al.,}
\titlerunning{A Hard Medium Survey with the ASCA GIS.IV}
\maketitle

\section{Introduction}

It is now largely accepted that X-ray obscured AGN play a significant (and
perhaps major) role  in the production of the Cosmic X-ray Background (CXB)
above 2 keV (Comastri et al., 1995; Gilli et al., 2001). These objects should be
the site where a large fraction  of the energy  density of the universe is
generated (e.g. Fabian and Iwasawa, 1999). 
In particular, the CXB synthesis models predict a large  density of high
luminosity (L$_x >10^{44}$ \es) X-ray obscured ($N_H > 10^{22}$ cm$^{-2}$)  
AGN which, according to the
Unified Model of AGN, should be characterized only by highly ionized narrow
emission lines in the optical domain  (the so called Type 2 QSO). Although
doubts are often cast on the existence of ``Type 2 QSO" 
(see e.g. Halpern, Turner \& George, 1999;
Akiyama et al., 2000), it is clear that  hard X--ray surveys, which are less
affected by the photoelectric absorption, should provide a fundamental tool
to detect and to study this elusive class of sources. 
In spite of these considerations, the number of high
luminosity, X-ray obscured AGN 
in spectroscopically  complete hard ($E > 2$ keV) X-ray
selected samples is still very low. 
For instance there are no type 2 QSO amongst the
34 sources of the  ASCA Large Sky Survey, which is 
a spectroscopically
identified and complete  sample in the 2-10 keV energy range (Akiyama
et al., 2000). A few type 2 QSO ``candidates" have been found using  ASCA
(Ohta et al., 1996; Akiyama, Ueda and Ohta, 2002) and ROSAT data  (Almaini et al., 1995;
Barcons et al., 1998; Georgantopoulos et al., 1999). Some 
of them have been recently  
discovered in {\it Chandra} and XMM  medium-deep surveys 
(Dawson et al., 2001; Crawford et al., 2002,
Stern et al., 2002; Norman et al., 2002;
Mainieri et al., 2002). 
Although these results seem to challenge the CXB synthesis model 
predictions, further X--ray spectroscopic investigations are needed.

According to the Unification Scheme of AGNs, a still unsettled
fraction of the radiation emitted by the central engine is absorbed by the
circumnuclear medium and re-emitted in the NIR/submillimeter bands. 
As a consequence, X-ray obscured AGN 
could also produce a significant  fraction of the
infrared background (Franceschini et al., 2002).
In this respect it is worth noting that Type 2 QSO candidates have been 
also selected in the infrared domain amongst the ultra- and hyper-luminous 
infrared galaxies 
(e.g. IRAS 09104+4109, Franceschini et al., 2000; Iwasawa, Fabian and Ettori, 2001). 
A detailed investigation of their Spectral Energy Distribution (SED) 
is thus mandatory to tackle many questions of the modern 
physical cosmology.

In this paper we discuss a Type 2 QSO, found in the ASCA Hard
Serendipitous Survey (HSS, Cagnoni, Della Ceca and Maccacaro, 1998; 
Della Ceca et al., 1999; Della Ceca et al., 2000a,b), 
for which we have  enough statistics to
perform a broad band (1 - 10 keV) X-ray spectral analysis. 
The object discussed here is also a strong radio source 
and it could have been classified as a Narrow Line Radio Galaxy, 
a class of ``potentially" obscured AGN (see Fabian, Crawford 
\& Iwasawa, 2002; Derry et al., 2003 for X-ray observations of 
powerful radio galaxies) well known from radio surveys 
and which has been 
studied up to $z=5.19$ (van Breugel et al., 1999; see 
also McCarthy 1993 for a review).
AX J0843+2942 is thus
representative of the radio-loud tail of the type 2 quasar population. 

The paper is organized as follows. In Section 2 we present the X-ray data
and their analysis. In Section 3 we report the optical
identification,  as well as the X-ray, optical 
and radio properties of AX J0843+2942.
Section 4 presents a discussion of the SED of  
AX J0843+2942
as well as the comparison  with the SED of Ultra Luminous 
Infrared Galaxies, other ``Type 2 QSOs", and ``normal" Radio-Quiet (RQ) and 
Radio-Loud (RL) QSOs 
from the literature. 
Finally, summary and conclusions are given in Section 5.  We
use $H_{0} = 50$ km s$^{-1}$ Mpc$^{-1}$ and $q_{0} = 0$ throughout.

\section {X-ray Data and Analysis}

AX J0843+2942 was discovered during the optical 
identification process of the X-ray sources in the ASCA 
(2-10 keV energy range) HSS.
The survey uses the data from the GIS2 instrument onboard ASCA 
(Tanaka, Inoue and Holt, 1994). 
The source attracted our attention because of its position in the 
hardness ratio diagram (cf. Della Ceca et al., 1999) indicative of a hard 
X-ray source.
The data preparation for the X-ray spectral analysis 
was performed as in Della Ceca et al. (2000a) and is summarized below.

To maximize statistics and signal-to-noise ratio, total counts (source +
background) were extracted from a circular region of 2$^{\prime}$ radius around
the source centroid in the GIS2 and GIS3 images. Background counts  were
taken from two circular, source-free regions of 4$^{\prime}$.75 radius close to
the source.  
GIS2 and GIS3 data, along with the relative
background files and  calibrations, were combined  following 
the recipe given in the ASCA Data Reduction Guide
\footnote{
see http://heasarc.gsfc.nasa.gov/docs/asca/abc/abc.html}.  
The combined GIS spectrum was rebinned in order to have 
a signal-to-noise ratio greater than 3 in each energy channel; 
the energy interval with useful data is from $\sim$ 0.9 to 
10 keV.
Unfortunately, there are no ASCA SIS  data associated to AX J0843+2942 since 
the source falls outside the SIS field of view.
Spectral analysis has been performed using XSPEC
11.01.  All the models discussed below have been filtered by the Galactic
absorption column density along the line of sight 
($N_{H_{Gal}} = 4.21\times 10^{20}$ cm$^{-2}$; Hartmann \& Burton, 1997). 
Unless explicitly quoted, all the errors on the fitted spectral parameters 
represent the 68\% confidence level for 1 interesting
parameter ($\Delta \chi^{2} = 1.0$).  
Basic information for AX J0843+2942 is given in Table 1. 

\begin{table*}
\begin{center}
\caption{Basic Informations on AX J0843+2942}
\begin{tabular}{llcrl}
\hline
\hline
ASCA Seq.    & RA; DEC (J2000) X       & Exp.(s)    & HR1              & Cts/s       \\
             & RA; DEC (J2000) Opt.    & S/N        & HR2              & Optical ID   \\
(1)          & (2)                     & (3)        & (4)              & (5)          \\
\hline
82011000     & 08 43 13; 29 42 46    & 76585      & $0.85 \pm 0.10$  & $(3.26\pm 0.25)\times 10^{-3}$  \\
             & 08 43 10; 29 44 05    & 10.2       & $0.14 \pm 0.07$  & Type 2 QSOs at z=0.398    \\
\hline
\hline
\end{tabular}
\end{center}
NOTE -- Columns are as follows: 
(1) ASCA Sequence number; 
(2) Source position as obtained from the centroid of the X-ray emission using the original astrometry from the ASCA GIS2 image
and, second line, the position of the optical counterpart of the X-ray source from the POSS II image;
(3) Exposure time (GIS2 + GIS3) of the observation and, second line, source Signal-to-Noise ratio in the 2-10 keV energy range from the ASCA HSS catalog;
(4) Hardness Ratio HR1 and, second line,  HR2 obtained using the measured counts in the $0.7-2$, $2-4$ and $4-10$ keV energy range (see Della Ceca 
et al., 1999);
(5) Net source count rate in the $0.9-10$ keV energy band and, second line,  
identification of the optical counterpart. The source count rate 
represents about 66\% of the total counts in the source extraction region. 
\end{table*}

\section{AX J0843+2942: a type 2 quasar candidate}

\subsection {X-ray Spectrum}

A single absorbed power-law model at the redshift of the  optical counterpart
(z=0.398; see Section 3.3) provides a good description  of the ASCA spectrum of
AX J0843+2942. The unfolded X-ray spectrum and the ratio between the
data and the  best fit model are shown in Figure 1. No emission lines or
absorption edges of statistical significance are  found superimposed on the
power law continuum. The 90\% upper limit on the Fe K$_{\alpha}$ line 
equivalent width at 6.4 keV (rest frame) is $\sim 400$ eV; this upper limit 
is consistent with the Fe K$_{\alpha}$  line equivalent width expected to be produced 
by trasmission in the 
same medium which absorbes the continuum (cfr. Lehly \& Creighton, 1993).  
Best fit spectral parameters, along with the observed
fluxes and intrinsic luminosities in the (0.5 -- 2) keV and (2 -- 10) keV
energy range, are reported in Table 2.

\begin{table*}
\begin{center}
\caption{Results of the spectral fit (GIS2 + GIS3 data) for AX J0843+2942 : absorbed power-law model at z=0.398}
\begin{tabular}{cccccc}
\hline
\hline
$\Gamma$             & $N_H$                      & Flux        &    Lx             &  $\chi^{2}_{\nu}$/dof  \\
                     & $10^{22}$ cm$^{-2}$        & $10^{-13}$ erg cm$^{-2}$ s$^{-1}$ & $10^{44}$ erg s$^{-1}$ & \\  
 (1)                 & (2)                        &  (3)        &   (4)             &  (5)                   \\
\hline
$1.72^{+0.3}_{-0.6}$ & $14.4^{+3.3}_{-5.2}$        & 0.09; 14.2  & 10.7; 18.9        & 0.59/15               \\
\hline
\hline
\end{tabular}
\end{center}
NOTE -- Allowed ranges are at 68\% confidence level for one interesting parameter ($\Delta \chi^{2} = 1.0$).
Columns are as follows: 
(1) Power-law photon index;
(2) Intrinsic absorbing column density;  
(3) Observed flux (de-absorbed from the Galactic $N_H$ value)
    in the 0.5-2 and 2-10 keV energy band;
(4) Intrinsic luminosity (i.e. the luminosity emitted from the nucleus) 
    in the 0.5-2 and 2-10 keV energy band;
(5) Reduced $\chi^{2}$ and degrees of freedom.
\end{table*}

In Figure 2 we show the confidence contours for the photon index ($\Gamma$) 
and the rest frame absorbing column density ($N_H$). 
While $\Gamma$ is not well
constrained (the 90\% confidence range  is between 0.7 and 2.6), the best fit
$N_H$ is $\simeq 1.44\times 10^{23}$
cm$^{-2}$, with a 90\%  confidence range between $6\times 10^{22}$ cm$^{-2}$
and  $2.4\times 10^{23}$ cm$^{-2}$, and a 99\% confidence level lower limit 
of $4\times 10^{22}$ cm$^{-2}$.
AX J0843+2942 is clearly  an absorbed
object; about 99.99\% and 50\% of the intrinsic rest-frame luminosity 
in the 0.5-2.0  keV and
2-10 keV energy band is absorbed. The intrinsic rest-frame luminosity 
($L_{(0.5 - 10) \rm keV} \simeq 3\times 10^{45}$ \es at z=0.398, see Section 3.3) 
is well within the ``High-Luminosity - QSO" regime.
The combination of high luminosity, high intrinsic
absorption and optical spectral properties (see Section 3.3) 
allow us to classify  AX J0843+2942 as an X-ray obscured ``Type 2 QSO". 

\begin{figure}[htb]
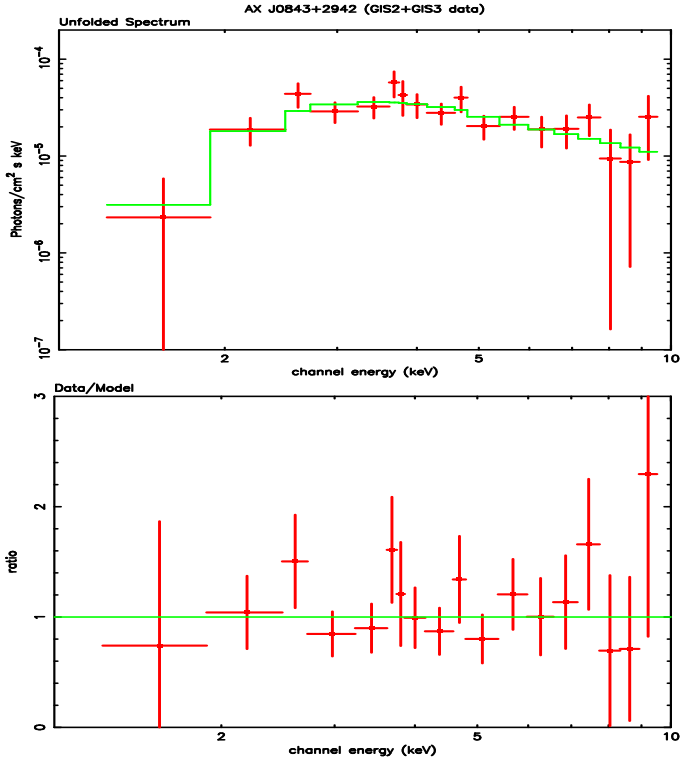

\parbox{10cm}{
\psfig{file=3656_f1a.ps,height=5.0cm,width=9cm,angle=-90} 
\psfig{file=3656_f1b.ps,height=5.0cm,width=9cm,angle=-90} }
\ \hspace{0cm} \
\caption[]{AX J0843+2942: unfolded X-ray spectrum (upper panel) and the 
ratio between the data (GIS2+GIS3) and the best fit 
absorbed power-law model (lower panel).
}
\end{figure}

\begin{figure}[htb]
\parbox{10cm}{
\psfig{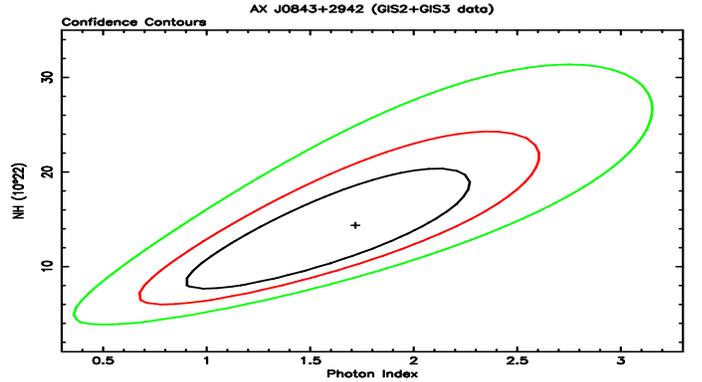} }
\ \hspace{0cm} \
\caption[]{Confidence contours (68\%, 90\% and 
99\% confidence level for two interesting parameters) for the photon index 
and the rest frame absorbing column density.}
\end{figure}

\subsection {Source Identification}

Figure 3 shows the Palomar Observatory Sky Survey II (POSS II) image centered
on the nominal X-ray position of AX J0843+2942 and its 90\% error circle of 2$^{\prime}$ 
radius. The cross marks the corrected  X-ray position of 
AX J0843+2942, obtained by applying to the source the same offset  
measured for the target of the ASCA observation. 
\footnote{The target
of this observation was MS0839.8+2938, an X-ray selected 
cluster of galaxies  at z=0.194. We have
determined the sky position of the bulk of its
X-ray emission using the ROSAT  HRI
observation rh800159.
Unfortunately AX J0843+2942 is on the 
border of this ROSAT HRI observation (at about 
16 arcmin offaxis) and is not detected.  
}

\begin{figure}[htb]
\parbox{10cm}{
\psfig{file=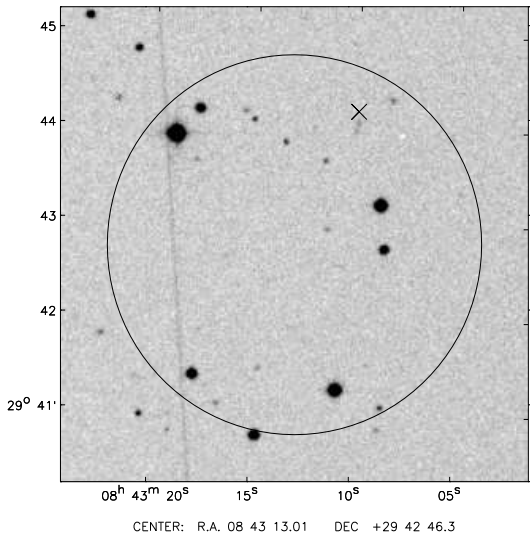,height=11.5cm,width=8.0cm,angle=0} }
\ \hspace{0cm} \
\vskip -1.5truecm
\caption[]{POSS II image ($5^{\prime} \times 5^{\prime}$) 
of the field centered at the ``raw" X-ray position of 
AX J0843+2942. The circle represents the  90\% confidence level error
box of 2$^{\prime}$ radius, while the  cross marks the position of 
the X-ray centroid when the same offset measured for the target of the 
observation (MS0839.8+2938) is applied.
}
\end{figure}

\begin{figure}[htb]
\parbox{10cm}{
\psfig{file=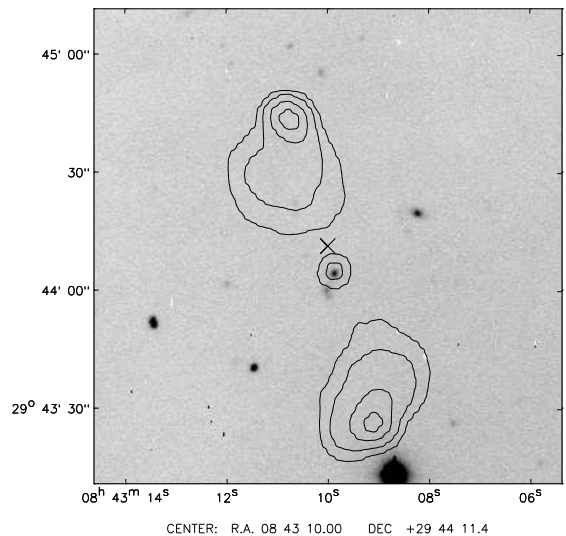,height=11.5cm,width=8.0cm,angle=0} }
\ \hspace{0cm} \
\vskip -1.5truecm
\caption[]{CCD image ($2^{\prime} \times 2^{\prime}$) 
centered at the revised ASCA position of
AX J0843+2942 with overlaid radio contours (at 1, 3, 10, 20 $\sigma$ 
above the background) from the FIRST radio survey.}
\end{figure}

In Figure 4 we show the optical image centered at the corrected ASCA position (cross)
with overlaid radio contours from the FIRST survey
\footnote{http://sundog.stsci.edu/}
(Becker et al., 1995; White et al., 1997).
The brightest optical object close to the 
X-ray position is positionally coincident with 
the core of a triple radio source having a total flux density (1.4 GHz) of 
$\sim 1$ Jy; 
the properties of this radio source are discussed in Section 3.4.
The probability to have a radio source with a flux 
$\sim 1$ Jy inside a circle of 
2$^{\prime}$ radius is $\sim 1.5\times 10^{-4}$ (Condon et al., 1998), which 
implies 0.03 spurious coincidences among the 188 ASCA HSS sources. 
Therefore we conclude 
that the X-ray and the radio emission are very likely to be 
related and are both due to the same object, i.e.  
the object positionally coincident with the core of the radio source 
(see Figure 4).
Optical, near infrared and far infrared 
photometry of this object are reported in Table 3.

\begin{table*}
\begin{center}
\caption{Photometry of AX J0843+2942}
\begin{tabular}{lcrlrrr}
\hline
\hline
                    &  Band        &  Frequency            &  Observed Flux Density  &  $\nu$ $f_{\nu}$       &   $\nu$ $L_{\nu}$ & Ref\\
                    &              &      Hz               &                         & erg cm$^{-2}$ s$^{-1}$ & erg s$^{-1}$  &    \\
\hline
Radio               &  178  MHz      &  $1.78\times 10^{8}$  &  $4.7$    Jy    &  $8.4 \times 10^{-15}$   & $7.6 \times 10^{42}$ &(1) \\
Radio (NVSS)        &  1.4  GHz      &  $1.4\times 10^{9}$   &  $985.7$ mJy    &  $1.4 \times 10^{-14}$   & $1.3 \times 10^{43}$ &(2) \\
Radio               &  4.85 GHz      &  $4.85\times 10^{9}$  &  $354$   mJy    &  $1.7 \times 10^{-14}$  & $1.5 \times 10^{43}$ &(3)  \\
Far-Infrared (IRAS) & 100$\mu$m      &  $3.0\times 10^{12}$  &  $<0.8$   Jy    &  $<2.4 \times 10^{-11}$ & $<9.1 \times 10^{45}$ &(4)  \\
Far-Infrared (IRAS) &  60$\mu$m      &  $5.0\times 10^{12}$  &  $<0.3$   Jy    &  $<1.5 \times 10^{-11}$ & $<5.8 \times 10^{45}$ &(4) \\
Far-Infrared (IRAS) &  25$\mu$m      &  $1.2\times 10^{13}$  &  $0.19$   Jy    &  $2.3 \times 10^{-11}$ & $8.7 \times 10^{45}$ &(4) \\
Far-Infrared (IRAS) &  12$\mu$m      &  $2.5\times 10^{13}$  &  $<0.1$   Jy    &  $<2.5 \times 10^{-11}$ & $<9.5 \times 10^{45}$&(4) \\
Infrared (2MASS)     &  K            &  $1.36\times 10^{14}$ &  $15.3$ mag   &  $6.9 \times 10^{-13}$   & $2.6 \times 10^{44}$ &(5)   \\
Infrared (2MASS)     &  H            &  $1.82\times 10^{14}$ &  $>16.3 $ mag   &  $<5.8  \times 10^{-13}$  &$<2.2 \times 10^{44}$ &(5)   \\
Infrared (2MASS)     &  J            &  $2.4\times 10^{14}$  &  $>16.5 $ mag   &  $<9.7 \times 10^{-13}$   &$<3.7 \times 10^{44}$ &(5)  \\
 Optical             &  E            &  $4.75\times 10^{14}$ &  $19.8  $ mag   &  $2.1 \times 10^{-13}$  & $8.1 \times 10^{43}$ &(6)    \\
 Optical             &  0            &  $7.4\times 10^{14}$  &   21      mag   &  $8.2 \times 10^{-13}$   &  $3.2 \times 10^{43}$&(6)    \\
X-ray (ASCA)        &  1.6 keV       &  $3.84\times 10^{17}$ &  $5.16\times 10^{-6}$  keV cm$^{-2}$ s$^{-1}$ keV$^{-1}$      
                                                                             &  $1.3 \times 10^{-14}$   &$1.3 \times 10^{43}$  &(7) \\
X-ray (ASCA)        &  5.1 keV      &   $1.22\times 10^{18}$ &  $1.3\times 10^{-4}$  keV cm$^{-2}$ s$^{-1}$ keV$^{-1}$      
                                                                             &  $1.1 \times 10^{-12}$  &$1.1 \times 10^{45}$  &(7)   \\
X-ray (ASCA)        &  9.2 keV       &  $2.21\times 10^{18}$ &  $1.0\times 10^{-4}$  keV cm$^{-2}$ s$^{-1}$ keV$^{-1}$      

                                                                             &  $1.5 \times 10^{-12}$  & $1.5 \times 10^{45}$ &(7)   \\
\hline
\hline
\end{tabular}
\end{center}
Note: The quoted luminosities have been K-corrected assuming the observed spectral indices in  
the radio ($\sim 0.78$), IR-optical ($\sim -1.8$) and X-ray  ($\sim 1.16$) domain.
References: (1) Pillington \& Scott., 1965;
            (2) total flux density from the NVSS survey (Condon et al., 1998);
	    (3) Gregory \& Condon, 1991;
	    (4) the upper limits derive from the IRAS Faint Source Catalog v2.0. The detection at 
	     25$\mu$m is at S/N=3 and derives from the ADDSCAN processing of IRAS data;
	    (5) from the 2MASS survey (see http://www.ipac.caltech.edu/2mass/);
	    (6) from the APM catalogue and corrected for blending;
	    (7) this work.

\end{table*}

\subsection {Optical Spectroscopy}

The optical object coincident with the 
core of the radio source was observed spectroscopically at
the 88'' telescope of the University of Hawaii  on 
February 12th, 2000. Two 20-minute exposures were
taken using the Wide Field Grism Spectrograph (WFGS) equipped
with the ``blue'' grism (400~grooves/mm) which gives a dispersion
of 4.1~\AA/pixel.  

The spectrum has been wavelength and  flux calibrated using a Hg-Cd-Zn
reference spectrum and  the photometric standard SAO98781, respectively. No
attempts have been made in order to have an absolute calibration  of the final
spectra. For the data reduction we have used the IRAF {\it longslit} package. 
The resulting spectrum, which is the sum of the two exposures, is presented in
Figure 5, while in Table 4 we report  optical line properties.

The widths (FWHM) of the permitted lines 
are all below $\sim $ 1200~km s$^{-1}$, 
suggesting a spectroscopic classification of the object
as a Narrow Emission Line object. The dominance of the 
[OIII]$\lambda$5007 when compared to the H$\beta$
([OIII]$\lambda$5007/H$\beta$=10) excludes the 
classification as starburst or HII-region galaxy. 
The  high [OIII]/H$\beta$ ratio and the lack of
evident Fe~II lines exclude also
the classification of AX J0843+2942 as Narrow Line
Seyfert~1 (NLS1) whose optical spectrum are characterized
by relatively weak forbidden lines 
([OIII]$\lambda$5007/H$\beta <$ 3) and strong Fe II 
multiplets (Osterbrock \& Pogge 1985). The analysis of the
H$\beta$ profile  does not reveal 
any strong evidence
of a broad wing underlying the narrow component, although
a better resolution spectrum is necessary to
exclude completely the presence of a broad H$\beta$ component. 

In conclusion, AX J0843+2942 is optically classified as Seyfert~2-like
object or, more appropriately, as an X-ray obscured type~2 QSO 
given its high X-ray luminosity and intrinsic absorption.

\begin{table*}
\begin{center}
\caption{Optical line properties of AXJ0843+2942}
\begin{tabular}{l r r r r}
\hline
\hline
Line   & Position  & z   & EW     & FWHM        \\
       &(observed) &     & (rest) &             \\
       & \AA       &     & \AA    & km s$^{-1}$ \\
\ (1)  & (2)       & (3) & (4)    & (5)         \\
\hline
\hline
$[$NeV$]$    & 4792 & 0.3987 &  80 & 1400 \\
$[$OII$]$    & 5211 & 0.3982 &  85 & 1050 \\
$[$NeIII$]$  & 5408 & 0.3978 &  90 & 1300  \\
H$\epsilon$  & 5549 & 0.3977 &  30 & 1200 \\
H$\delta$& $\sim$5729 & 0.397: & $\sim$14&$-$ \\
H$\gamma$+$[$OIII$]$& $\sim$6070/6102 & 0.399: & $\sim$85 & $-$ \\
HeII     & $\sim$6548 & 0.397: &  $\sim$ 11 & $\sim$1200 \\
H$\beta$ & 6798 & 0.3985 & 30 & 850 \\
$[$OIII$]$   & 6934 & 0.3983 & 90 & 750 \\
$[$OIII$]$   & 7001 & 0.3982 & 300 & 750 \\
\hline
\hline
\end{tabular}
\end{center}
\label{lines}
\end{table*}

\begin{figure*}[htb]
\parbox{10cm}{
\psfig{file=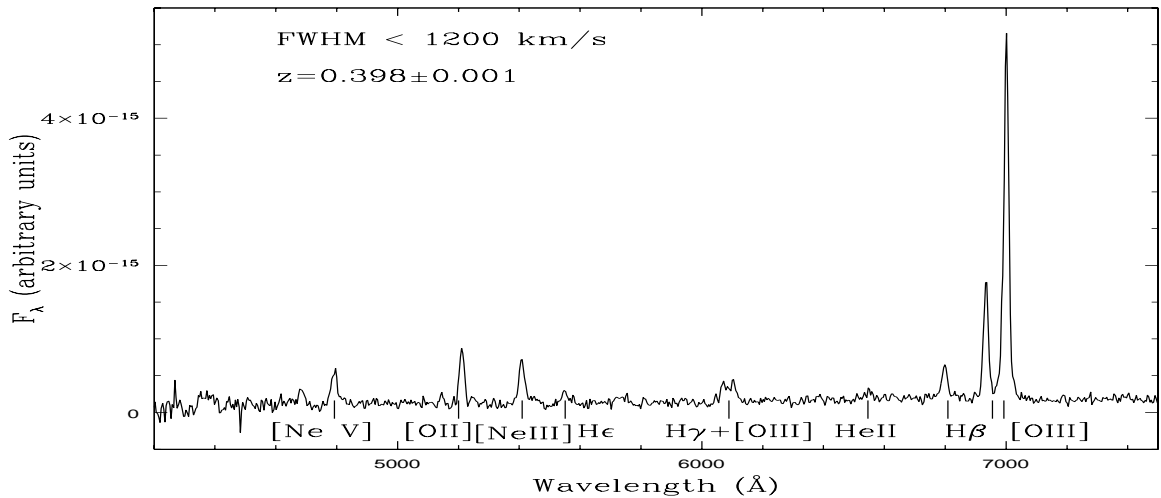,height=7.0cm,width=16.2cm,angle=-90} }
\ \hspace{0cm} \
\caption[]{Optical spectrum of AX J0843+2942 with the prominent emission 
features labeled
}
\end{figure*}

\subsection{Radio Properties}

The radio counterpart of AX J0843+2942 is a triple radio source 
having the typical Fanaroff-Riley type II morphology, i.e. 
sharp edged lobes and  bright hot spots (see Figure 4) .
The absolute optical magnitude 
($M_E \simeq -21.5$) and the total radio power 
($\sim 9 \times 10^{33}$ \es Hz$^{-1}$, see below) 
are consistent with the FRII classification 
according to the dividing line in the $M - L_{radio}$ plane
between FRI and FRII radio galaxies 
(Ghisellini and Celotti, 2001 and references therein).

The total 1.4 GHz radio flux from the FIRST
survey is $878\pm 44$  mJy corresponding to a
radio power of  $(8.0\pm 0.4) \times 10^{33}$ erg s$^{-1}$ Hz$^{-1}$
at the redshift of the source; 
the core flux density (power) is $20.9\pm 1.1$  mJy 
($P_{1.4 GHz} = 0.19\pm 0.01 \times 10^{33}$ erg s$^{-1}$ Hz$^{-1}$).
The source is also
detected in the lower spatial resolution 
NVSS survey
\footnote{http://info.cv.nrao.edu/$\sim$jcondon/nvss.html}
 (Condon et al., 1998)
as a double radio source.  The total NVSS flux density (radio power) is
$986\pm 40$ mJy ($9.0\pm 0.4 \times 10^{33}$ erg s$^{-1}$ Hz$^{-1}$). 
The radio flux density from the NVSS is a more reliable measure of the 
total flux since the VLA-B configuration used for the FIRST survey could 
miss some of  the diffuse extended emission. 
This radio source has been also detected  at 178 MHz (4C +29.31,
Pilkington \& Scott, 1965) and at 4.85 GHz (Gregory \& Condon, 1991).
Assuming a power-law model the radio spectral index between 178 MHz and 
4.85 GHz is $\sim 0.78$, consistent with the typical radio spectral index 
of lobe-dominated AGN.
Total radio flux densities and powers are reported in Table 3.
The projected separation of the two bright hot spots 
is $\sim 80$ arcsec on sky, corresponding to a physical projected  size of
0.57 Mpc at the redshift of the source;
the ratio between the core flux and the total flux is about 0.021.
Given the discussed radio properties, the object is clearly a radio-loud 
and lobe-dominated AGN.  In this respect 
we note that the measured size is comparable with the size of the 
known giant radio galaxies studied by Lara et al. (2001) and 
Schoenmakers et al. (2001).

\section {Discussion}

It is  interesting to compare the SED
 of AX J0843+2942 with the SED of
other classes of AGNs to see if there are similarities and/or differences.
In particular we are interested in the comparison with the classes of AGNs 
where absorption effects can strongly modify their appearance and 
classification depending on the observational wavelength.
Photometric data used to construct the SED of AX J0843+2942 have been 
summarized in Table 3.

Particular care must be taken on 
how to normalize the {\it observed} SEDs
since many of the objects discussed below are either 
affected by heavy absorption 
(which modifies the intrinsic SEDs) or  
contaminated (and even dominated) by the possible starburst contribution,
particularly in the infrared domain.
Indeed there is now strong evidence that the
processes of star-formation and AGN emission mostly happen in a  high-density
medium ($N_H > 10^{23-24}$ cm$^{-2}$), characterized by high dust  extinction
of the UV-optical flux and strong photoelectric absorption of the  soft
X-rays (see e.g. Levenson et al., 2001; Iwasawa 1999 and reference therein).

The first comparison (Figure 6) is with three well known local AGN: the
ultraluminous  infrared galaxy (ULIRG hereafter)  
NGC 6240 ($L_{FIR} \sim 5\times 10^{45}$ \es at z=0.0245), 
the nearby far-infrared galaxy  
NGC 4945 ($L_{FIR} \sim 3\times 10^{44}$ \es at z=0.00187)
and the powerful radio-galaxy Cygnus A 
($L_{FIR} \sim 2\times 10^{45}$ \es at z=0.0562).

\begin{figure*}
\begin{center}
\begin{tabular}{cc}
\includegraphics[width=0.50\textwidth]{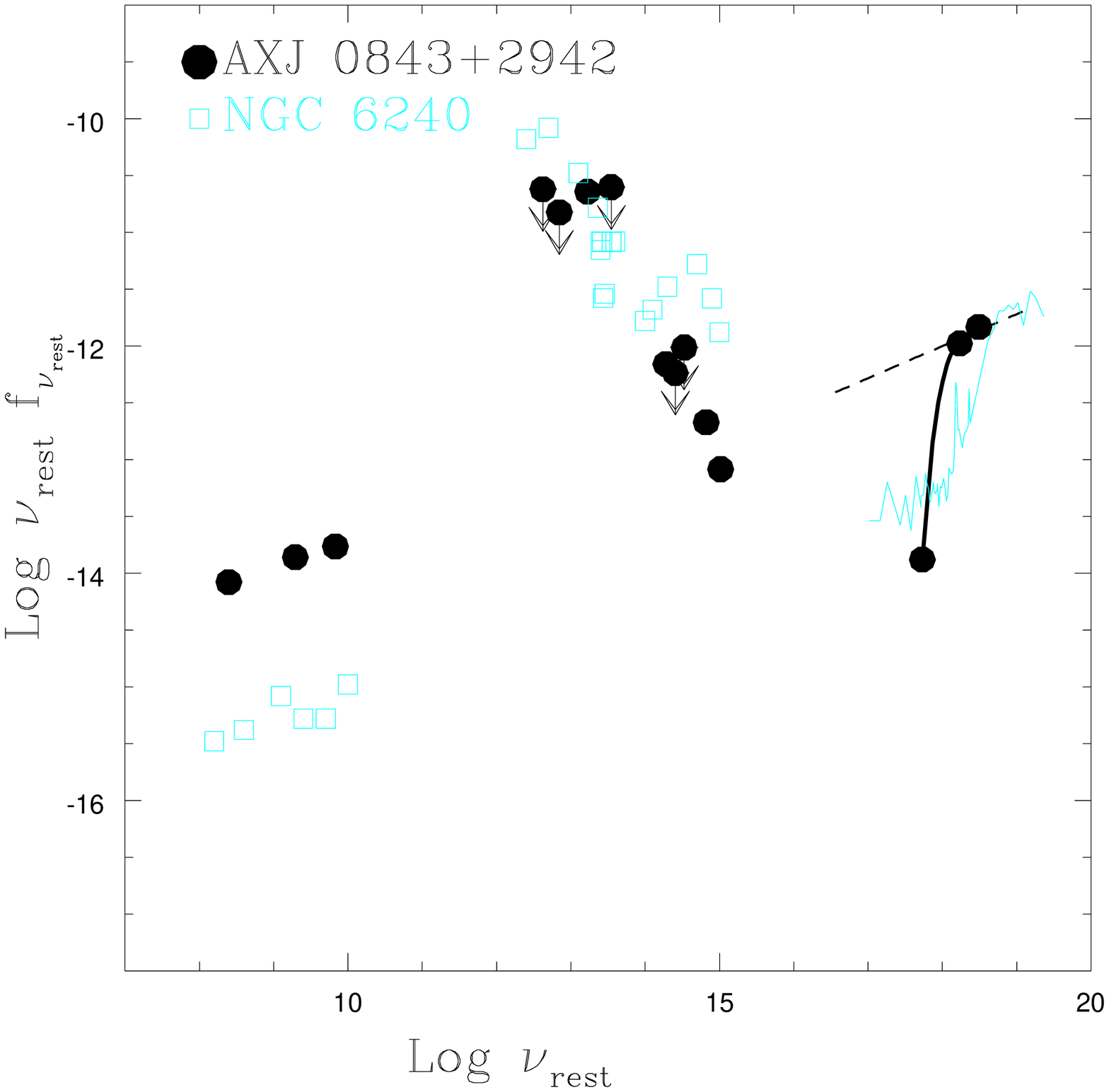}&\includegraphics[width=0.50\textwidth]{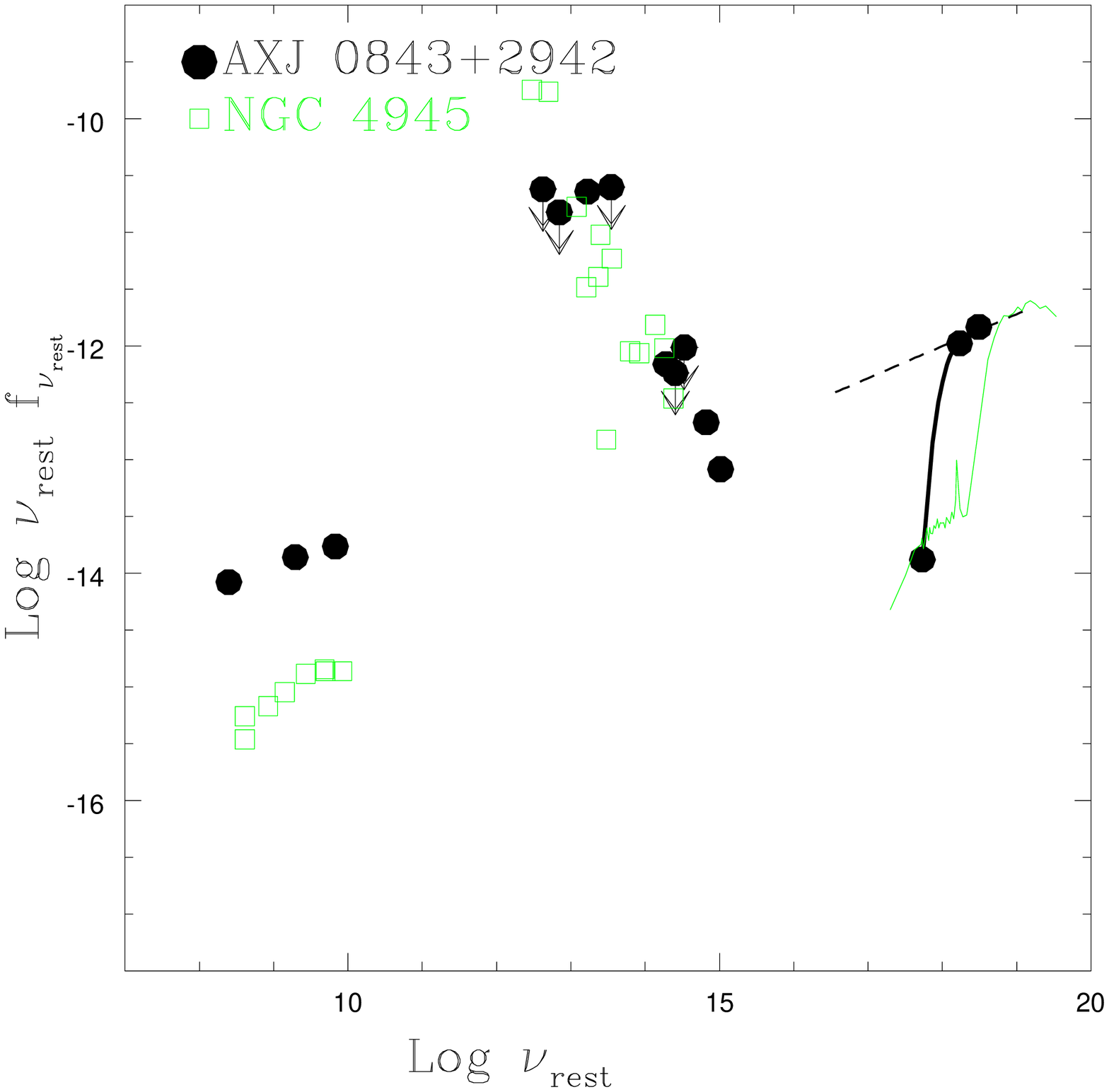}\\
\includegraphics[width=0.50\textwidth]{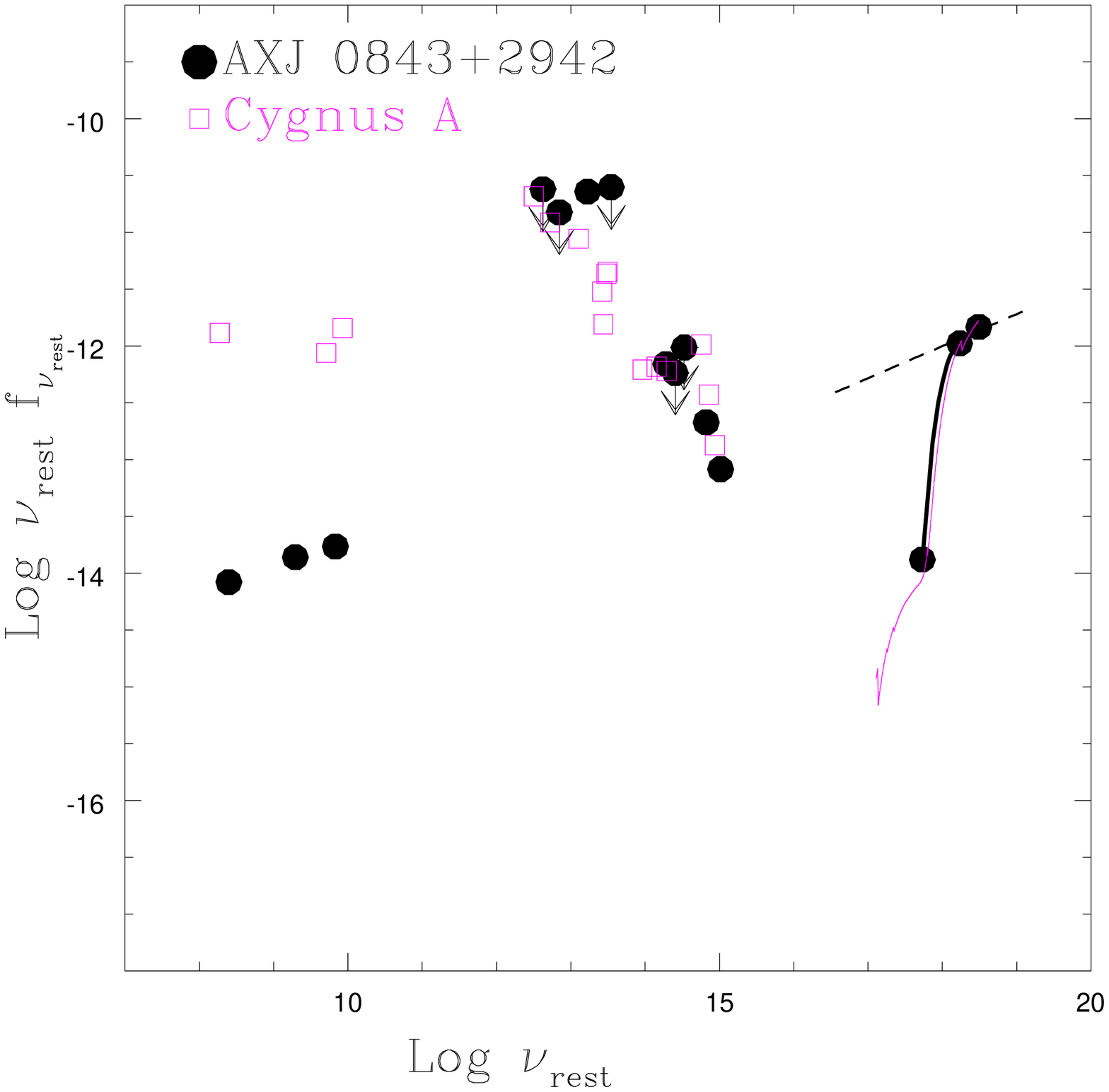}&
\begin{minipage}{9cm}
\vspace*{-8cm}
\caption{Comparison between the SED of AX J0843+2942 (filled circles, thick
solid line and  arrows at around $2 \times 10^{14}$ Hz) and the SED of   
a) NGC 6240 (panel a: open squares and thin solid line, adapted from Iwasawa et al., 2001), 
b) NGC 4945 (panel b: open squares and thin solid line, adapted from Iwasawa et al., 2001) 
and c) Cygnus A (panel c: open squares and thin solid line, data from NED with the X-ray 
spectral model from Young et al., 2002). 
Photometric data from radio to optical wavelengths for 
AX J0843+2942 are reported in Table 3; in the X-ray domain we  have plotted the
best fit spectral model.  The dashed line at $\nu_{rest} \simeq 10^{16.5} -
10^{19}$ Hz$^{-1}$  represents the intrinsic ``unabsorbed" spectra of
AX J0843+2942. 
The flux scale on the Y axis refers to  AX J0843+2942, while the X axis reports
the rest frame frequency. 
}
\end{minipage}\\
\end{tabular}
\end{center}
\end{figure*}

NGC 6240 
\footnote {
A recent {\it Chandra} observation of NGC 6240 has revealed the presence of
two active galactic nuclei of comparable luminosity in the core of this object
(Komossa et al., 2003). In the following discussion  we will consider 
the integrated emission from the two nuclei.} 
and NGC 4945 have been classified as LINER and/or starburst galaxies
on the basis of  optical (Veilleux~et~al.~1995)
and mid-/far-IR (Genzel~et~al.~1998) spectroscopy.
For both objects, however, the {\it Beppo}SAX PDS observations at $E>10\:$keV have 
clearly revealed the presence of a deeply buried AGN 
($N_H \simeq$ few $\times 10^{24}\:$ cm$^{-2}$)
with a QSO-like intrinsic luminosity in the case of 
NGC~6240 ($L_{2-10 \rm keV} \sim 3.6\times 10^{44}\:$\es; Vignati~et~al.~1999) 
and a Seyfert-like luminosity in the case of NGC~4945  
($L_{2-10 \rm kev} \sim 4\times 10^{42}\:$\es; Guainazzi~et~al.~2000).
Cygnus A is a nearby and powerful radio galaxy having a compact 
nucleus which is well described, in the X-ray domain, by an absorbed 
($\sim 2 \times 10^{23}$ cm$^{-2}$) power law spectrum. Its
unabsorbed luminosity ($L_{(2-10 \rm keV)} \sim 3.7\times 10^{44}$ erg s$^{-1}$)
is in the QSOs range (Young et al., 2002).

The comparison of these three objects with  AX J0843+2942  is
shown in Figure 6a (NGC 6240), in Figure 6b (NGC 4945) and in  Figure 6c (Cygnus A).
The SED of NGC 6240, NGC 4945 and Cygnus A have
been tied to that of AX J0843+2942 under the hypothesis that  the nuclear and
intrinsic AGN emission emerges above $\sim 20$ keV
\footnote{It is worth noting that when a column density exceeds $10^{24}$
cm$^{-2}$, even the high-energy continuum  (i.e., above the absorption cut-off)
will be suppressed significantly by Compton down-scattering (see Matt et al.
1999). This is a geometry-dependent effect; the suppression is larger as the
covering fraction of the absorber is smaller. Therefore the intrinsic power-law
continuum could be somewhat higher (e.g., 20-50 per cent for $N_H \simeq$
$10^{24} - 5\times 10^{24}$ cm$^{-2}$) in Compton-Thick objects, NGC6240,
NGC4945, and IRAS 09104+4109 (see Figure 7a for this latter object).}
.

\begin{figure*}
\begin{center}
\begin{tabular}{cc}
\includegraphics[width=0.50\textwidth]{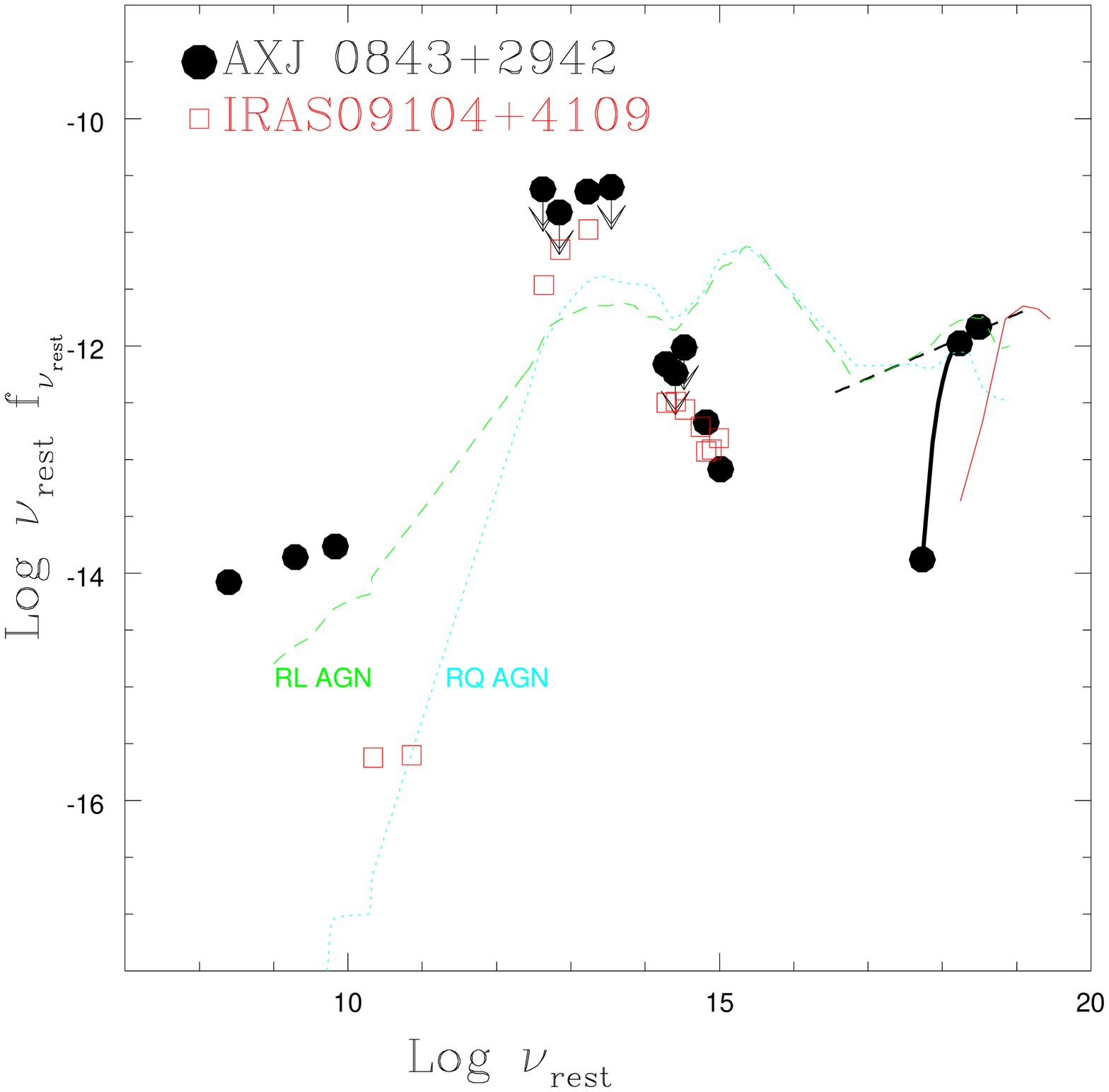}&\includegraphics[width=0.50\textwidth]{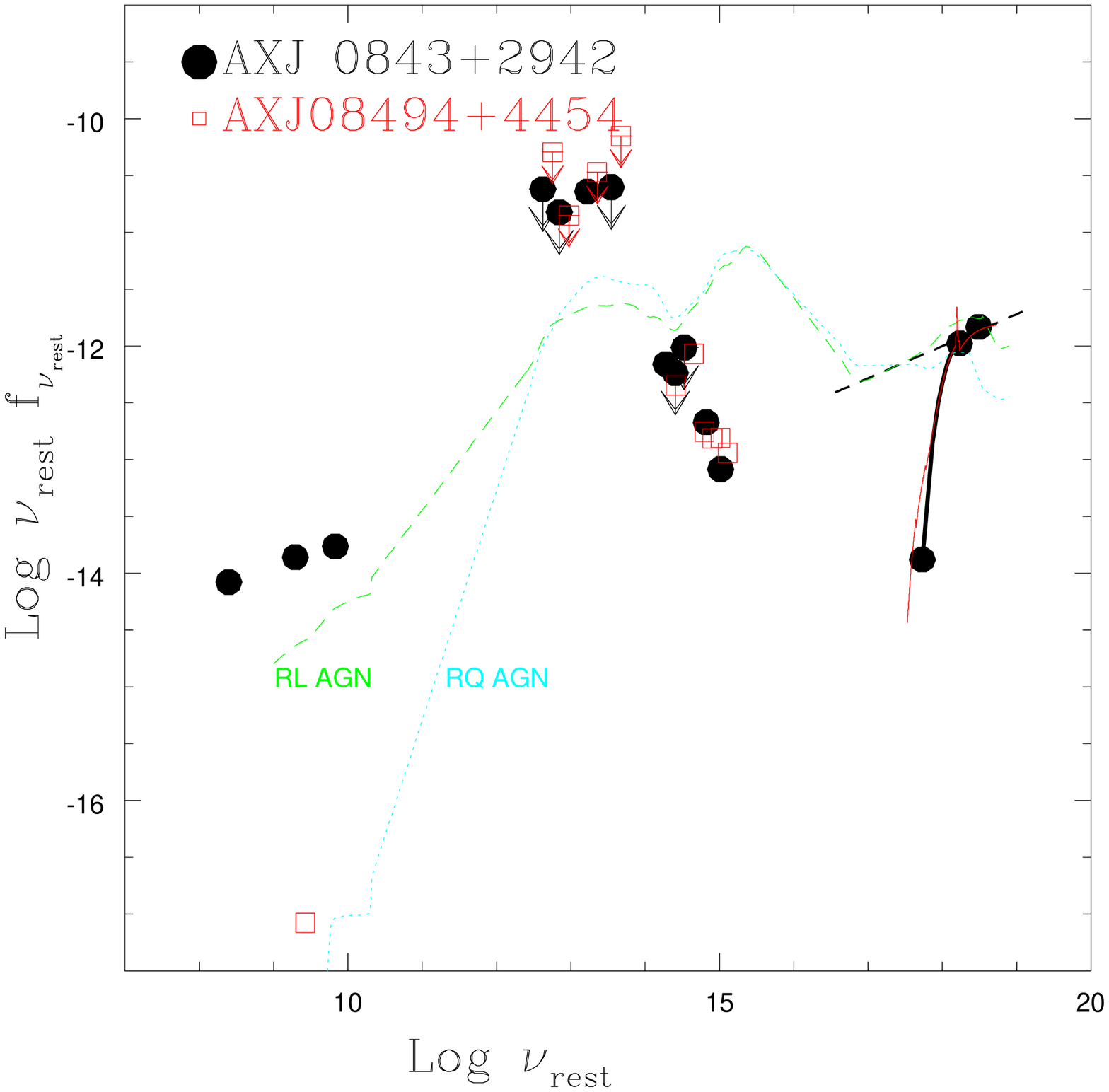}\\
\end{tabular}
\caption{Comparison between the SED of AX J0843+2942 (filled circles, thick
solid line and  arrows at around $2 \times 10^{14}$ Hz), the Type 1 RL and  RQ
QSOs (as labelled in the figure), the Type 2 QSOs  IRAS09104+4109 (panel a:
open squares and thin solid line) and the Type 2 QSOs AX J08494+4454.  
(panel b: open squares and thin solid line). The dashed line at $\nu_{rest}
\simeq 10^{16.5} - 10^{19}$ Hz$^{-1}$  represents the intrinsic ``unabsorbed"
spectrum of AX J0843+2942. The SED of Type 1 RL and RQ QSOs have been adapted
from Elvis et al., 1994;  the SED of IRAS09104+4109 from Franceschini et al.,
2000; and the  SED of AX J08494+4454 from Akiyama et al., 2002 (and reference 
therein). The flux scale on the Y axis refers to  AX J0843+2942, while the X
axis refers to the rest frame frequency.  
}
\end{center}
\end{figure*}

The second comparison  (shown in Figure 7a, and 7b) is with the SED of  distant 
type 1 and type 2 AGN.
The SED of ``normal" type 1 RL and RQ QSOs (labeled in Figure 7a,b) are taken
from Elvis et al.,  1994.  
As for the distant type 2 QSOs we have used two objects 
for which enough photometric data are available 
to investigate their broad band properties over a large energy range  
($\sim$ 10 decades in frequency) and a {\bf firm} measure of the 
intrinsic  AGN spectra above $\sim$ 20 keV is also available.
The two objects are IRAS09104+4109 and AX J08494+4454.

IRAS09104+4109, the most luminous object in the $z<0.5$ Universe, 
is an ULIRG ($L_{IR} \sim 2\times 10^{46}$ \es) at z=0.442.
{\it Beppo}SAX observations (Franceschini et al., 2000; see also 
Iwasawa, Fabian and Ettori, 2001 for the {\it Chandra} observation) 
have allowed the investigation of this source up to $\sim 80$ keV, and 
have clearly 
revealed the presence of a deeply buried AGN. 
The absorbing column density towards the nucleus is 
$\geq 5\times 10^{24}$ cm$^{-2}$  and the unabsorbed (2-10) keV X-ray 
luminosity is $\sim 8 \times 10^{45}$ erg s$^{-1}$, which is within the 
range of quasar luminosities.

AX J08494+4454 is a type 2 QSOs candidate at $z=0.9$ found in the course of  the
optical identification of ASCA deep surveys in the Lynx field (Ohta et al., 
1996).  The 0.5 - 10 keV {\it Chandra} spectrum is hard and is well described 
by a power law continuum absorbed by an hydrogen column density of  $\sim 2
\times 10^{23}$ cm$^{-2}$; the unabsorbed luminosity in the  2-10 keV energy
range is $\sim 7 \times 10^{44}$ \es (Akiyama, Ueda and Ohta,  2002). Recent
deep Subaru/IRCS J-band spectroscopic observations seem to suggest the presence
of a broad $H{\alpha}$ component at the bottom of the  narrow $H{\alpha}$
emission line, implying that AX J08494+4454 could be a  luminous cousin of
Seyfert 1.9 objects (Akiyama, Ueda and Ohta,  2002).

The SED of Type 1 RL and RQ
QSOs have been normalized to the  intrinsic ``unabsorbed" flux of AX J0843+2942
at $\sim 2$ keV. The SED of  IRAS09104+4109 and AX J08494+4454  have been tied to
that of AX J0843+2942 under the hypothesis that their emission above $\sim 20$
keV represents the intrinsic AGN emission. 

As expected, both the local and the distant absorbed AGN in figure 6 and in figure 7
lack the optical-UV bump which characterizes  the ``normal" Type 1 QSOs.
However they show  a very similar behavior in the
optical-infrared regime despite their different X-ray absorbing  column density
towards the nucleus (ranging from  $N_H \sim 2\times 10^{23}$ to $> 5 \times 10^{24}$ cm$^{-2}$).
The only object which deviates significantly is NGC 6240 in the optical
and near-IR bands; for this object the optical near-IR photometric points were taken from 
NED and are usually measured with large apertures (K. Iwasawa, private communication).
Two additional objects  (CDF-S 202: a type 2 QSO at z=3.7 from Norman et al., 2002; 
and CXO52: a type 2 QSO at z=3.288 from Stern et al., 2002) show a similar
behaviour although it is difficult to evaluate the real
``intrinsic"  spectrum given the poor X-ray statistics.

We tried to reproduce the near-infrared and optical colors of  
AX J0843+2942 assuming a ``typical" QSO spectra template (continuum plus
emission lines, adapted from Elvis  et al., 1994 and Francis et al., 1991) and
the extinction curve from  Cardelli et al. (1989). A good description of the
infrared-optical colors  of AX J0843+2942 is obtained assuming $E_{B-V} \sim
0.6$. 
Provided that the X-ray and optical/NIR emission come from the same region 
(which may not be necessarily true), 
the implied $E_{B-V}/N_{H}$ is $\sim 4 \times 10^{-24}$ mag cm$^2$, a
factor about 40 less than the Galactic standard 
value of $\sim 1.7 \times 10^{-22}$ mag cm$^2$ 
(Bohlin et al., 1978).  This is
a well know problem since the {\it Einstein} era  (Maccacaro et al., 1982) and
has been recently discussed by  Maiolino et al. (2001) in the context of different
dust properties in the circumnuclear region of AGNs. 

Finally under the simple assumptions used by Norman et al. (2002;  
e.g. the black hole is generating 
X-rays with an efficiency of $\sim 1\%$ relative to the Eddington luminosity) 
we estimate a black hole mass of $\sim 2 \times 10^{9}$ M$_{\odot}$.
This mass is similar to that measured in other steep spectrum radio quasars 
having a similar total radio power of AX J0843+2942 (Lacy et al., 2001).   
By estimating the nuclear radiative output (L$_{ion}$) using the well
established  relation between the luminosity in narrow emission lines
(believed  to result from the photoinization by the nuclear accreting
radiation) and the  radio power, we obtain (cfr. Ghisellini \& Celotti, 2001)
$L_{ion} \sim 4\times 10^{46}$ \es.  This value implies $L_{ion}/L_{EDD}
\sim 0.13$ and an accretion rate  relative to the Eddington accretion rate of
1.3 (assuming an efficiency of  0.1). These values are consistent with the
statement of Ghisellini \& Celotti (2001) that, for the same black hole mass,
the FRII radio-galaxies have a higher accretion  rate if compared with FRI type
objects.

\section {Summary and Conclusion}

We have presented the X-ray, optical and radio properties of  AX J0843+2942, a
high luminosity Radio-Loud Type 2 AGN found in the ASCA Hard Serendipitous Survey. This
source, positionally coincident with the  core of a triple and strong ($S_{1.4
GHz} \sim 1$ Jy;  $P_{1.4 GHz} \sim 9\times 10^{33}$ erg s$^{-1}$ Hz$^{-1}$) 
radio source,  is spectroscopically identified with a Narrow Line  object
(intrinsic FWHM of all the observed emission lines $\lae 1200$ Km s$^{-1}$)
at z=0.398, with line features and  ratios typical of Seyfert-2 like
objects.

The X-ray spectrum is best described by 
an absorbed power-law model with 
photon index of $\Gamma$ = $1.72^{+0.3}_{-0.6}$ and intrinsic 
absorbing column density of  
$N_H$ = $1.44^{+0.33}_{-0.52} \times 10^{23}$ cm$^{-2}$ 
(99\% confidence level lower limit 
of $4\times 10^{22}$ cm$^{-2}$). 
The intrinsic luminosity in the 0.5 -- 10 keV energy band is   
$\simeq 3\times 10^{45}$ \es, well within the range of 
quasar luminosities.

The high X-ray luminosity, coupled with the  high intrinsic
absorption, optical spectral properties 
and radio power allow us to
classify AX J0843+2942 as an X-ray obscured Radio-Loud ``Type 2 QSO".

We find strong similarities in the SED of AX J0843+2942 and the 
SED of local absorbed AGNs and distant Type 2 QSO in the optical
near-infrared regime, despite the very different X-ray
absorbing  column densities towards the nucleus.

The near-infrared and optical colors of  
AX J0843+2942 can be reproduced assuming a ``typical" QSO spectrum template, 
$E_{B-V} \sim 0.6$, implying an $E_{B-V}/N_{H}$ that is a factor 
40 less than the Galactic standard value.

The estimated black hole mass ($\sim 2 \times 10^{9}$ M$_{\odot}$) 
is consistent with a relatively high accretion rate 
to power the bolometric luminosity of the QSO, in agreement with the
dividing line between FRI and FRII type objects proposed by 
Ghisellini \& Celotti (2001).

\begin{acknowledgements}

We thank K. Iwasawa, M. Akiyama and A. Franceschini for providing the
data,  shown in figure 6 and 7, in a tabular form. 
This work received partial financial support from the
Italian Ministry for University and Research (MURST)  and from 
ASI (I/R/037/01).
K.Iwasawa is also thanked for useful comments in the referee report.  
This research has made use of the NASA/IPAC
extragalactic database (NED), which is operated by the Jet Propulsion
Laboratory, Caltech, under contract with the National Aeronautics and
Space Administration.  We thank all the members of the ASCA team who
operates  the satellite and maintains the software data analysis and the
archive.

\end{acknowledgements}

\end{document}